# High-temperature dust formation in carbon-rich astrophysical environments

Guy Libourel ●[1] ✉, Marwane Mokhtari[2], Vandad-Julien Rohani[3], Bernard Bourdon ●[2], Clément Ganino ●[4], Eric Lagadec[1], Philippe Vennéguès[5], Vincent Guigoz[5], François Cauneau ●[3] & Laurent Fulcheri ●[3]

Condensation processes, which are responsible for the main chemical differences between gas and solids in the Galaxy, are the major mechanisms that control the cycle of dust from evolved stars to planetary systems. However, they are still poorly understood, mainly because the thermodynamics and kinetic models of nucleation or grain growth lack experimental data. To bridge this gap, we used a large-volume three-phase alternating-current plasma torch to obtain a full high-temperature condensation sequence at an elevated carbon-to-oxygen ratio from a fluxed chondritic gas composition. We show that the crystallized suites of carbides, silicides, nitrides, sulfides, oxides and silicates and the bulk composition of the condensates are properly modelled by a kinetically inhibited condensation scenario controlled by gas flow. This validates the thermodynamic predictions of the condensation sequence at a high carbon-to-oxygen ratio. On this basis and using appropriate optical properties, we also demonstrate the influence of pressure on dust chemistry as well as the low probability of forming and detecting iron silicides in asymptotic giant branch C-rich circumstellar environments as well as in our chondritic meteorites. By demonstrating the potential of predicting dust mineralogy in these environments, this approach holds high promise for quantitatively characterizing dust composition and formation in diverse astrophysical settings.

Cosmic dust is of crucial importance for the evolution of our Galaxy because it actively participates in the cycle of matter (gas and dust) from the interstellar medium (ISM) to stars and back from stars to the ISM[1]. Dust grains form principally in gas outflowing evolved stars and in supernova explosion ejecta. Among the progenitors, asymptotic giant branch (AGB) stars are assumed to be the major contributors to the global cosmic dust budget[2]. During their late evolutionary stages, low- and intermediate-mass stars (<8 $M_\odot$) become red giants, increasing their radius by two to three orders of magnitude and decreasing their surface temperature to 2,000–3,000 K. Further evolution of these stars (AGB phase) leads to a important mass loss of gaseous molecules and dust grains, forming circumstellar envelopes (CSEs). The mass loss rates of AGB stars, determined using various observational methods, are typically in the range of $10^{-8}$ to $10^{-5}$ $M_\odot$ year$^{-1}$, but values as high as $10^{-4}$ $M_\odot$ year$^{-1}$ have been found for more extreme objects[3]. Around cool stars, dust is formed in dense shells created by stellar pulsations or large-scale convective motions that propagate gas beyond the stellar surface[4], where gas cooling yields dust condensation. These dust grains are then accelerated away from the star by radiation pressure, dragging the gas away through friction, creating outflows with typical velocities of 5–30 km s$^{-1}$ and eventually feeding the ISM[5], thus contributing to the material out of which our Solar System formed.

---

[1]Université Côte d'Azur, Observatoire de la Côte d'Azur, CNRS, Laboratoire Lagrange, Nice, France. [2]Laboratoire de Géologie de Lyon Terre Planètes Environnement, Ecole Normale Supérieure de Lyon, CNRS, Université Lyon I, Lyon, France. [3]PSL Research University, MINES ParisTech, PERSEE – Centre Procédés, Énergies renouvelables et Systèmes énergétiques, Valbonne, France. [4]Université Côte d'Azur, Observatoire de la Côte d'Azur, CNRS, Laboratoire Géoazur, Valbonne, France. [5]Université Côte d'Azur, CNRS, CRHEA-Centre de Recherche sur l'Hétéro-Epitaxie et ses Applications, Valbonne, France. ✉e-mail: libou@oca.eu

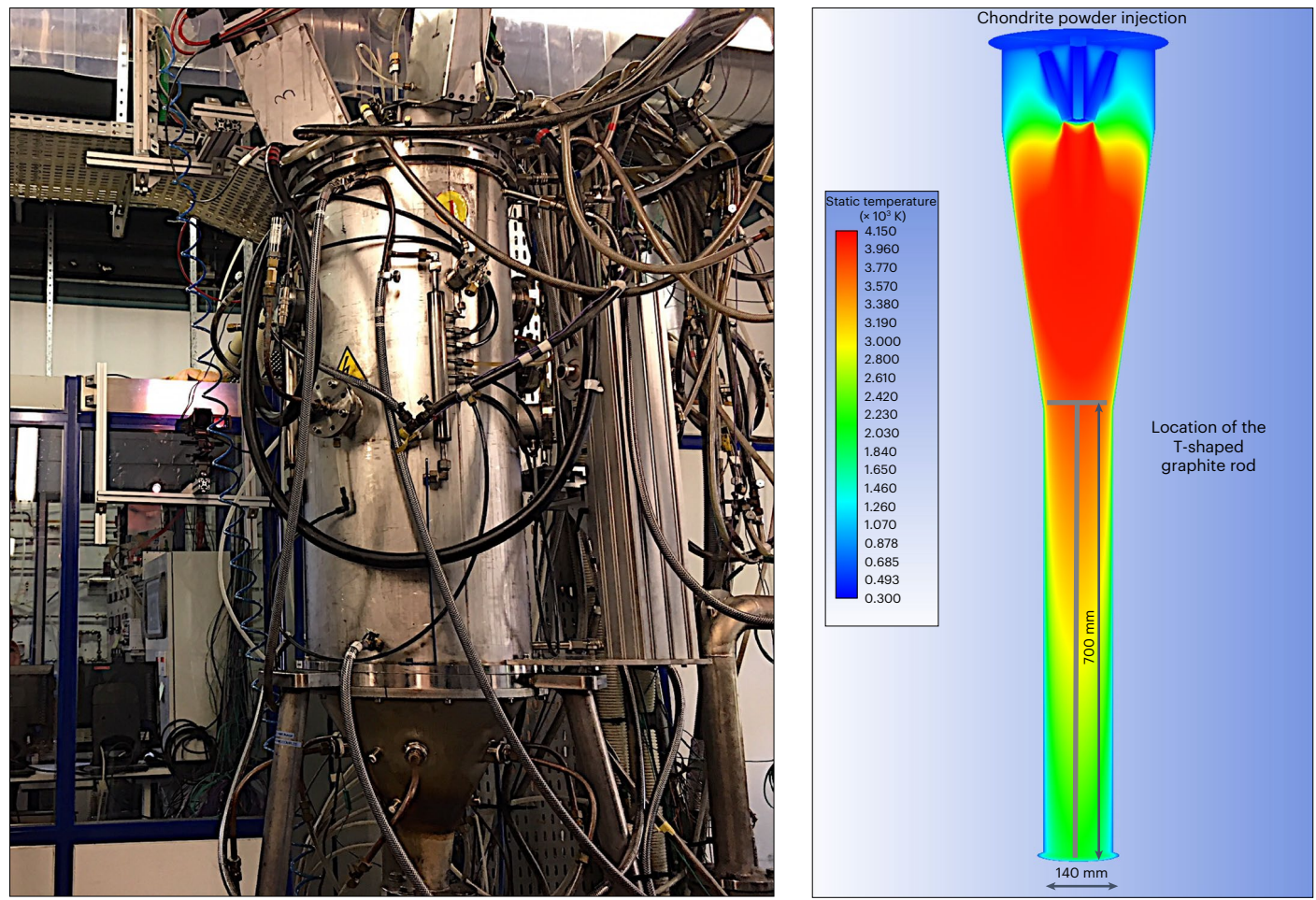

**Fig. 1 | Experimental device and thermal regime in the condensation chamber. a**, A 240-kW three-electrode arc plasma torch (PERSEE, Mines ParisTech, Sofia Antipolis, Nice, France) in which 600 g of a 20 µm-sized powder of NWA 869 ordinary chondrite was processed for an hour. **b**, Simulated 3D thermal regime inside the condensation chamber using the ANSYS Fluent computational fluid dynamics software. A removable T-shaped graphite rod was placed at the centre of the chamber and recovered after the experiment for chemical and mineralogical analyses of the condensates. A 3D simulation was performed in a computational domain that did not explicitly incorporate the T-shaped graphite rod. To account for the presence of the rod, the torch power was set to the experimentally determined value, maintaining the appropriate source temperature while ensuring an overall energy balance. To achieve this, the losses induced by the rod were redistributed onto the walls of the chamber, preserving the global thermal behaviour of the gas flow. Although the thermal distribution may not be locally exact near the rod location, the rod temperature can be considered to lie within the thermal boundary layer that forms near the walls. Equivalence calculations suggest that the rod temperature is between approximately 1/8 and 1/6 of the nozzle diameter away from the walls.

Dust forms from matter present during star formation and from the elements produced by the stars themselves. The dust composition changes as the star evolves, notably when the carbon-to-oxygen (C/O) ratio changes during the episodic third dredge-up process[6]. Thermodynamic equilibrium calculations have provided convincing explanations for the marked chemical differentiation between oxygen-rich and carbon-rich AGB stars based on the high bond energy of carbon (that is, 1,077 kJ mol$^{-1}$ ≈ 11.1 eV (ref. 7)), and that CO is a very stable gaseous molecule at relatively high temperatures. Consequently, the dust chemistry in the inner CSE of the AGB stars, $R = 1 - 10R_*$, where $R_*$ is the stellar radius, is primarily controlled by the C/O ratio. When C/O ≤ 1 (O-rich stars), all carbon is bound to the CO molecule, and excess oxygen participates in oxygen-rich dust formation[8]. In contrast, when C/O ≥ 1 (C-rich stars), carbon is in excess, whereas all the oxygen is bound to CO, and hydrocarbon molecules and carbon-rich grains are formed[9,10]. In addition to C/O, the temperature and pressure within the inner CSEs obviously play a substantial role (see below, 'Condensation in C-rich stellar atmospheres').

Complex dynamic effects in stellar outflows[3] and highly diverse circumstellar O-rich and C-rich grain inventories in chondritic meteorites[11,12] highlight, however, the need to consider beyond simple thermodynamic equilibrium to understand dust formation around AGB stars, including kinetics, nucleation/seeding processes and models such as chaotic solid formation[8,13–15]. This issue has been the subject of extensive theoretical studies[13]. Reference 16 recently re-emphasized the importance of kinetics for dust formation in astrophysical environments and the need for laboratory condensation experiments to provide insights into the formation of dust and the reactions that allow their growth. While the nucleation and growth of simple metal or H$_2$O grains from a gas have been extensively studied for applications in atmospheric or material science[17–21], the condensation of more complex gases, such as those of stellar outflows, is less well studied. The difficulty lies in simulating astrophysical conditions in the laboratory: (1) producing a multi-elemental refractory gas of controlled composition and (2) condensing it in equilibrium/disequilibrium conditions at a very high temperature[22–25].

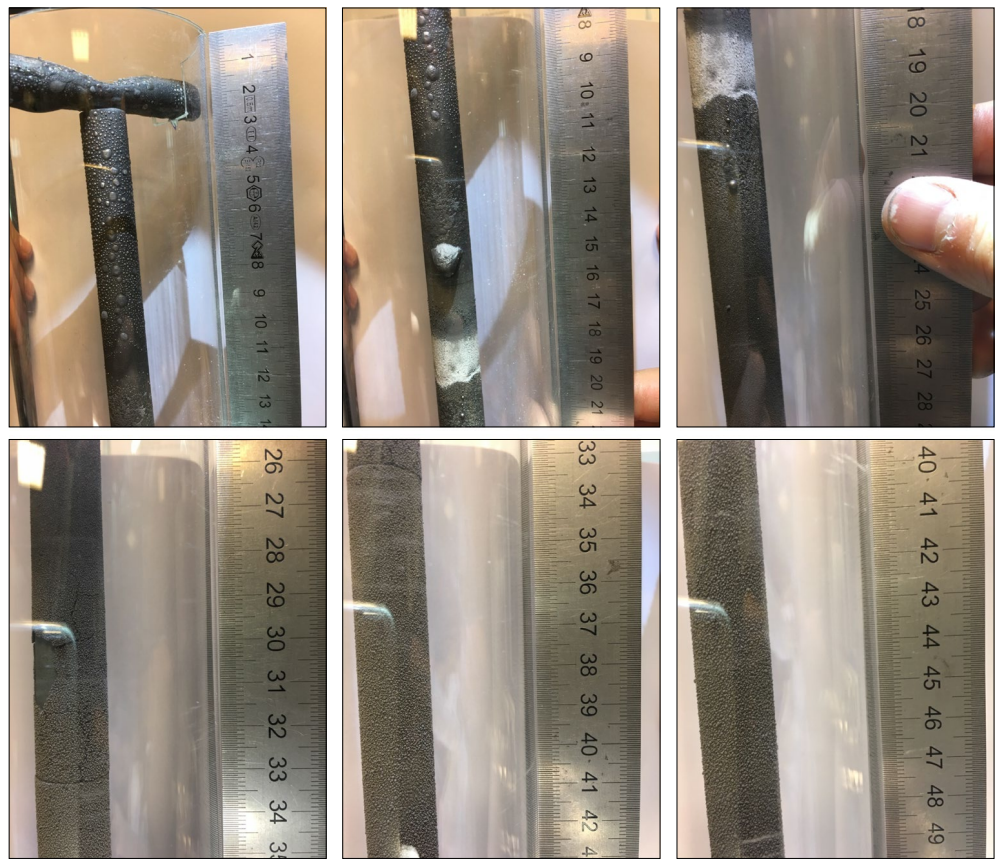

**Fig. 2 | Photographs of condensates on the T-shaped graphite rod after the experiment.** Note the change in the structure (for example, wetting and flaking) and grain size of the condensates from the top (high temperature) to the bottom (low temperature). See text.

In this study, 600 g of a 20 μm-sized powder of Northwest Africa (NWA) 869 ordinary chondrite was processed for an hour inside a large-volume 250-kW three-phase alternating-current plasma torch with hot graphite electrodes operating at the Mines ParisTech-PERSEE in Sophia Antipolis, Nice, France (Fig. 1). Chondritic dust particles were injected using argon gas for propulsion into a 5,000 K $H_2$–Ar admixture gas plasma formed between three graphite electrodes (Methods). The vapour formed in the plasma flowed downwards in a cylindrical metre-sized graphite-coated chamber, in which condensation occurred according to the thermal gradient at a pressure of $P \approx 10^5$ Pa. Nitrogen was injected during the run to avoid the flammability of hydrogen. The experiment was terminated by shutting down both the electrical power and powder supply. A removable T-shaped graphite rod was placed at the centre of the chamber and recovered after the experiment for chemical and mineralogical analyses of the condensates (Fig. 1 and Extended Data Fig. 1). Even if this experiment remains far from CSE conditions, for example, $P \ll 10^2$ Pa, which are technically impossible to achieve at present, we show in the following how such high-temperature condensates help us put stringent constraints on dust formation in stellar outflows.

## Results

Owing to this large-volume device, the recovery of several hundred grams of high-temperature condensates obtained at a high C/O ratio is a milestone. The three-dimensional (3D) thermal regime inside the chamber, as well as the trajectory and velocity of fictive particles (Fig. 1 and Extended Data Fig. 1), were obtained using the ANSYS Fluent computational fluid dynamics software (Methods; ref. 26). While condensation occurred in the whole condensation chamber, the T-shaped graphite rod holder offered enough material (Fig. 2) to detail the chemistry (Extended Data Table 1) and mineralogy (Extended Data Table 2) of the condensates between 2,400 and 1,300 K at ambient pressure and C/O ratio ≈ 0.93 (Extended Data Fig. 2). The high C/O ratio was regulated by the consumption of graphite from the three electrodes during the experiment (see below, Methods 'Experimental Approach').

X-ray diffraction pattern and secondary electron– and back-scattered electron–scanning electron microscopy characterization revealed that the condensates obtained at the highest temperatures, that is, >2,050 K, were dominated by iron silicides ($Fe_xSi_y$ with $x/y$ atomic ratio ≈ 1.2). They formed millimetre-sized droplets that wet the T-shaped graphite rod (Fig. 2). Well-crystallized SiC was bathed in silicides together with rarer and smaller crystals of $Al_4SiC_4$ and TiC (Fig. 3 and Supplementary Fig. 1). A temperature of about 2,050 K marked the onset of condensation of the oxygen-bearing and sulfur-bearing phases with the occurrence of calcium aluminate (silica-poor) glass and of CaS–MgS oldhamite, respectively. Calcium aluminates that are initially molten wet the immiscible FeSi molten droplets and SiC. Oldhamite condensed first as a Ca-rich phase and shifted to a MgS-rich composition as the temperature decreased. AlN (some of which are whiskers) and TiN nitrides appear below on the T-rod in a narrow interval of temperature between 2,000 and 1,900 K. In the same range of temperature, olivine $Mg_2SiO_4$ started to crystallize from the melt in close association with FeSi and SiC. Below a temperature range of 1,850–1,800 K, graphite became the dominant phase, whereas carbides were no longer detectable. Low-Ca pyroxene $MgSiO_3$ started to condense below 1,700 K, forming remarkable one-dimensional whiskers by an FeSi catalyst-assisted vapour–liquid–solid growth process (Supplementary Fig. 2). Iron silicides, $Fe_xSi_y$, have $x/y$ atomic ratios close to 2 in these low-temperature condensates. Crystals of periclase have also been found in the lowest-temperature part of the very fine-grained condensates.

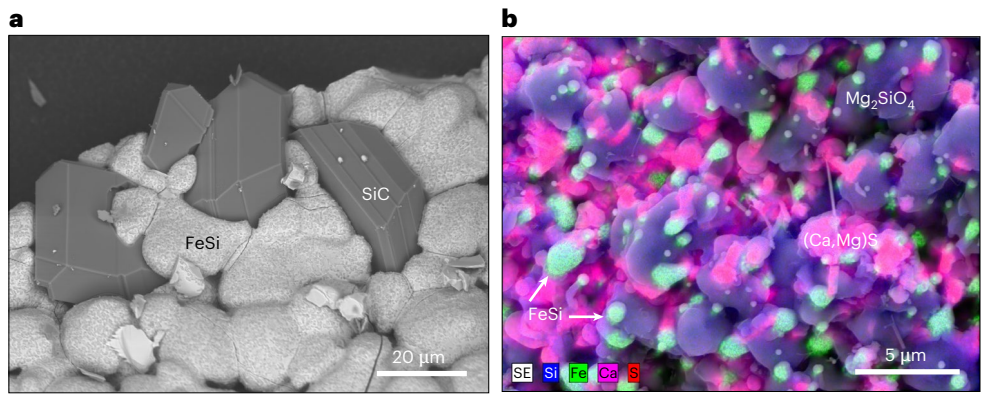

**Fig. 3 | Representative scanning electron microscope images of high-temperature condensates obtained at a high C/O ratio. a**, Well-crystallized SiC bathed in molten FeSi (now solid) from sample 13 condensed at a temperature of around 2,150 K. **b**, Overlapped X-ray elemental map in Ca (pink), Fe (green), Si (blue) and S Kα (red) and secondary electron map of no. 8 condensates obtained at a temperature of 1,700 K showing an association of oldhamite (CaS–MgS), Mg-rich olivine ($Mg_2SiO_4$) and molten droplets of iron silicide (FeSi).

Extended Data Tables 1 and 3 show the bulk chemistry of both the starting NWA 869 chondrite composition and the 14 aliquots of condensates sampled on the T-shaped graphite rod holder measured by inductively coupled plasma optical emission spectrometry. As condensation proceeded downwards in the chamber, condensates were first enriched in Fe, Si, Ti and C, and then became enriched in Al, Ca and S, and next in Mg, a change in chemistry that is in good agreement with their mineralogy (Fig. 4). However, sulfur was enriched in the low-temperature condensates (<1,600 K). Potassium was not found in the coldest condensates, whereas sodium was present only in trace amounts.

## Discussion

### Laboratory condensation of chondritic gas at high C/O ratios

Our results show a sequence of carbides, sulfides, nitrides and oxides that is generally consistent with calculations made in previous studies that correspond to the low-pressure conditions of protoplanetary disks or stellar atmospheres at high C/O ratios[14,27–30]. However, there are some differences due to the duration and high-pressure conditions ($\approx 10^5$ Pa) of the experiment. Carbon in our experimental condensates was mainly in the form of carbides at high temperatures (above 1,850 K) and transitioned to graphite at lower temperatures. No evidence of amorphous carbon was observed. The most important difference was the stability of liquids under these conditions, namely iron silicide metallic liquid and oxide/silicate liquid, which are immiscible. The large SiC crystals that crystallized from the FeSi liquid[31] and the various one-dimensional crystals/whiskers of AlN and $MgSiO_3$ catalysed by liquid FeSi droplets on their tips provided convincing examples of the vapour–liquid–solid process and its importance under high-pressure condensation conditions (Fig. 3 and Supplementary Figs. 2 and 3).

Regarding the chemistry of condensates (Extended Data Table 1), the calculated compositions based on thermodynamic equilibrium or fractional condensation did not match the measured compositions, whereas the calculated mineralogical sequences did (Methods). This suggests that an equilibrium condensation sequence, that is, the order of phase appearance during condensation, is not a good discriminant for identifying the exact process of condensation. Thus, we explored a partial fractional condensation model (Extended Data Fig. 3), where only a fraction $f$ of what was expected to condense in equilibrium could effectively condense due to kinetic limitations (Methods).

The condensation pattern of Si, Fe, Al and Ca determined using the model fits well with the experimental data (Fig. 4). The Fe concentrations in the condensates remained high (~60%) until 1,950 K when phases other than FeSi started condensing (SiC, AlN, CaS and oxide melt). The modelled $x/y$ ratio of the $Fe_xSi_y$ molten alloy ranged from 1 to about 2 as the temperature decreased, which is perfectly consistent with the measured compositions. The Si concentration remained almost constant along the rod, showing roughly constant condensation of this element in various phases: FeSi, SiC, oxide melt, followed by silicates. The Ca concentration showed a sharp increase until CaS became stable (Fig. 4). Its concentration increased until it reached a maximum at 1,900 K. At this temperature, CaS was no longer stable and Ca was mainly partitioned in the oxide melt. Al concentrations showed a rather similar pattern, with a sharp increase when AlN became stable, followed by a short decrease until it reached a plateau at 1,900 K when AlN was no longer stable, and Al condensed in the oxide melt. Both Ca and Al concentrations sharply decreased after 1,800 K when the Ca-poor and Al-poor phases started condensing (that is, olivine and periclase).

Despite the overall good agreement, some discrepancies between the model and the experiment can be observed (Fig. 4). The main issue is related to sulfur, which has very low predicted concentrations in the low-temperature sections compared with the measured concentrations. The main S-bearing phase in the thermodynamic simulations was CaS, whereas the experimental results were not consistent with the calculated condensation curve for this species. To obtain a CaS phase, the concentration of Ca should have been much higher than what was observed. Furthermore, no new sulfide phase was identified in the low-temperature segments of the rod with X-ray diffraction. Thus, there is a clear problem with the sulfur data that could have arisen for several reasons, including (1) analytical artefact with the measurement, (2) contamination by the condensation chamber, (3) thermodynamic under-estimation of the amount of S that could be dissolved in the FeSi and/or (4) notable solubility of sulfur in silicate liquids under reducing conditions[32]. Despite this difference in concentrations, the predicted and observed patterns of S concentrations are very similar, with the first peak of concentration at 1,950 K corresponding to the initial stability field of CaS and a sharp increase until a plateau around 1,750 K corresponding to the second stability field of CaS.

Our model, which incorporates a kinetically limited condensation mechanism, effectively reproduced the experimental results for most of the major elements at high C/O ratios. This provides experimental validation of the thermodynamic calculations employed to predict condensation sequences under such conditions. Notably, it was also established that the kinetically inhibited condensation sequence does not differ from the equilibrium sequence, whereas the relative abundances of the condensed phases and the resulting bulk composition do differ. Consequently, the bulk composition, rather than the mineralogical sequence of condensation, serves as a more reliable indicator of the extent of kinetic limitations during the condensation process.

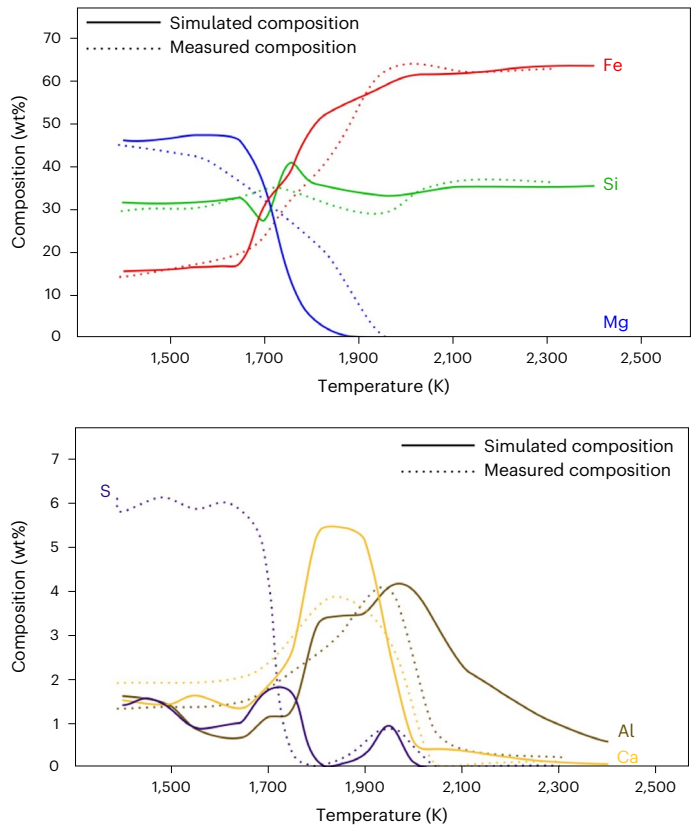

**Fig. 4 | Chemical composition of condensates.** Kinetically limited condensation with various f values for the phases is considered (see text). The solid lines correspond to the modelled compositions, and the dashed lines correspond to the measured compositions.

It is crucial to highlight that while the traditional distinction between O-rich and C-rich dust chemistry commonly aligns with a C/O value of unity, the C/O ratio derived from our C-rich dust-generating experiment drops below this threshold, settling at C/O = 0.93. Thus, this reported ratio is typical of an S-type star rather than a C-type star. According to previous thermodynamic calculations[28], the transition between reduced phases (for example, carbides and nitrides) and oxygen-bearing phases (for example, oxides and silicates) was also observed at a C/O ratio of approximately 0.96 at a total pressure of $10^2$ Pa, indicating a notable pressure influence on this transition. As shown in Fig. 5, our calculations confirmed a pressure-dependent shift in the C/O ratio required for oxide formation. As the pressure approached $10^5$ Pa, the C/O threshold dropped well below ≈1, indicating a strong destablization of oxide phases in favour of carbides. This is concomitant with the dramatic increase in the proportion of carbon–hydrogen bonds as the pressure increases, as indicated by the increase in the $C_2H_2$/CO ratio of the vapour phase (Fig. 5). The observed preference for C–H bond formation can be attributed to the inherent weakness of this bond compared to the C–O bond in carbon monoxide. The C–H bond energy is several hundreds of kilojoule per mole lower, making it easier to break and participate in the formation of new reactive carbon species in the vapour phase under pressure. This pressure effect can be rationalized by a reaction involving the main gas species of the type:

$$2\,CO_{(g)} + 3\,H_{2(g)} = C_2H_{2(g)} + 2\,H_2O_{(g)}$$

which, according to Le Chatelier's principle, will be shifted towards the product side ($C_2H_2$ formation) as the pressure increases ($P\Delta V$ negative), where $V$ is volume and $\Delta V = V_{\text{Products}} - V_{\text{Reactants}}$. This aligns perfectly with the observed increasing stability of the C-bearing condensed phases at higher pressures and with the presented experimental data (Fig. 5). In light of this, our condensation sequence at $10^5$ Pa with C/O = 0.93 replicated the sequence observed in CSE with higher C/O ratios.

### Condensation in C-rich stellar atmospheres

The evolution of AGB stars is controlled by two main processes: dredge-ups that bring material from the inner shell to the outer layers, thus increasing the C/O ratio of the stellar atmosphere[33,34], and mass loss, including C through stellar winds[35]. Dust formation by condensation is crucial because stellar winds are driven by momentum transfer from the photons emitted by the star to the surrounding dust[3,36]. Dust grains, accelerated by radiative pressure, drag the surrounding gas, creating stellar winds. The efficiency of this process depends on dust opacity. Winds form only when the radiative pressure overcomes the star gravity, requiring a dust opacity exceeding a critical value determined by the star mass and luminosity. However, momentum transfer is not the only light–matter interaction that plays an essential role. In particular, the light absorption properties of some minerals can heat up the grains, such that the actual temperature of the dust grains can be different from the local gas temperature. Thus, these grains can be heavily heated, and their temperature can exceed their condensation temperature. Consequently, their condensation can be inhibited, destabilizing these minerals, even though the gas temperature should allow condensation. This phenomenon has already been described for Fe-bearing olivines around AGB M stars[37]. This effect contributes to the increase in the effective oversaturation of this phase. As the gas and dust flow away from the star, the temperature and pressure decrease such that, at some point, the predicted dust temperature corresponds to its condensation temperature[38,39]. The evolution of AGB stars critically depends on the properties of the condensing grains with respect to light–matter interactions (Methods and Extended Data Table 4).

The condensation temperatures of dust with a C-star composition were then calculated using the thermodynamic database used to model our experimental condensation sequence. The starting bulk composition was solar abundance[40] with an increased C abundance. The distance at which a mineral first condensed was calculated for the temperature and pressure profiles in the extended atmosphere of a representative AGB star[14]. In this calculation, the abundance of C was increased over the solar value starting from a value of 0.93 (corresponding to our experiment) up to 1.5, because most C-rich AGB stars have 1 < C/O < 1.5. The results of these calculations are plotted in Fig. 6, where dust temperatures are also reported. In principle, the minimum distance at which dust can condensate is located at the intersection of the condensation and dust temperature curves. For TiC, SiC and olivine, the distance is close to the star (<2 $R_*$). Under these conditions, the pressure is high enough (>$10^{-2}$ Pa) for condensation to occur. However, for CaS, FeSi and C, condensation can only occur at distances greater than 5 $R_*$. At this distance, the pressure is lower than $10^{-5}$ Pa, and condensation must take place at a much lower rate owing to a lower local density. To further determine whether the formation of FeSi dust could take place at such a large distance from the star, we evaluated the timescale necessary for FeSi dust growth compared with the timescale of the outflow itself using the approach described in ref. 41 (Methods). The results are shown in Extended Data Fig. 4. The growth timescale for dust particles between $10^{-1}$ and $10^{-3}$ μm is always an order of magnitude longer than the outflow timescale. This shows that the provided dust temperatures prevent condensation at a close distance from the star (here, >5 $R_*$), and the low density in this environment then becomes prohibitive for allowing sufficiently rapid grain growth. Consequently, our kinetic models predicted that FeSi is unlikely to form in the outflows of C-rich AGB stars, despite thermodynamic predictions favouring its condensation under these conditions. In addition, one could note that detection of the infrared signature for FeSi in stellar outflows is itself difficult[42] due to the lack of distinct spectral features.

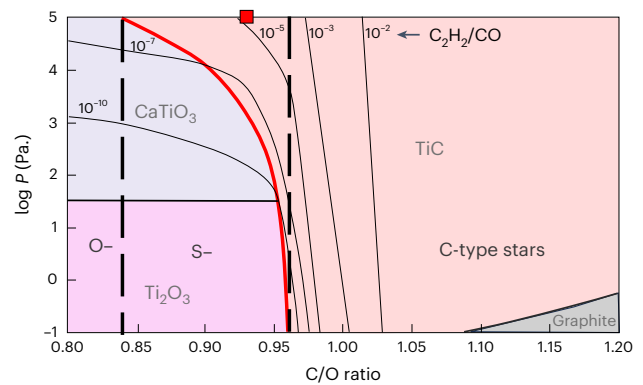

**Fig. 5 | Influence of pressure on dust chemistry.** Pressure versus C/O ratio diagram showing the stability field of the first condensates using the starting bulk composition of the solar composition from ref. 40, except for the C abundance. In this range of $P$ and C/O ratios, $Ti_2O_3$ and $CaTiO_3$ were stable at low C/O ratios; note that $Al_2O_3$ (not shown) was stable at lower C/O ratios[28]. At a higher C/O ratio, the oxides were no longer stable, and TiC and graphite were becoming predominant. The red square corresponds to our experimental conditions. The thin lines represent the $C_2H_2$/CO ratio in the gas phase at the saturation of the first condensates, that is, $Ti_2O_3$, $CaTiO_3$, TiC or graphite, depending on the C/O ratio and pressure. Notice the notable pressure influence on the transition between the carbon-bearing and oxygen-bearing phases (red curve). As the pressure approached $10^5$ Pa, the C/O threshold dropped well below ≈1, indicating a strong destabilization of oxide phases in favour of carbides (see text).

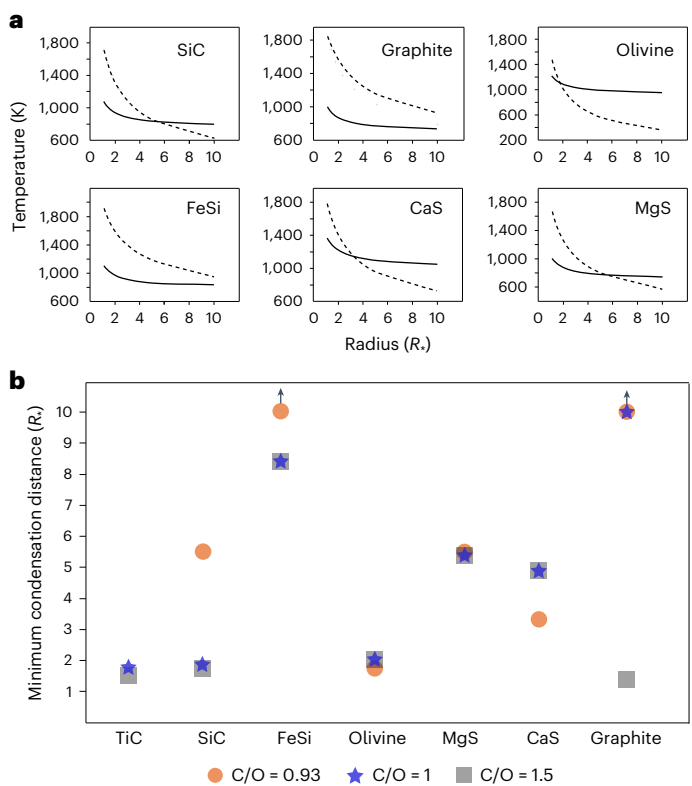

**Fig. 6 | Probability of forming and detecting dust in AGB circumstellar environments. a**, Comparison of the thermodynamic condensation temperature (solid line) and dust temperature (dashed line) as a function of the distance from an AGB star with a C/O ratio of 0.93 for six different minerals. The composition of the gas and the pressure profile are from ref. 14. The surface temperature of the star is considered to be 2,500 K. Condensation of minerals can only occur when the actual dust temperature is lower than the condensation temperature, that is, the minimum condensation distance is the intersection of the dashed and solid lines. **b**, Condensation distances for various mineral phases calculated for C/O ratios of 0.93, 1 and 1.5. The distances were expressed as multiples of the star radius and were obtained using the same approach as in **a**.

Assuming that our calculations captured the essence of processing affecting the dust generated in C-rich AGB stars, these results can be compared with direct observations of AGB stellar outflows and presolar grains preserved in primitive chondrites. A mid-infrared survey of evolved stars revealed a clear dichotomy in the chemical properties of dusty outflows. Carbon dust, be it amorphous carbon or crystallized graphite[10], and silicon carbide appear together in the outflows of C-enriched stars, whereas silicates, alumina and water ice dominate the dust emission of stars with C/O « 1 (ref. 43). C-rich dust, such as graphite, SiC and polycyclic aromatic hydrocarbons (PAHs), shows emission features at 3–30 μm. Carriers of the 21 and 30 μm features in the outflows of C-rich stars are currently still debated[44]. For instance, the feature observed at around 30 μm in the outflows of C-stars is often attributed to MgS, but its formation is unclear[44–46]. An important feature is that the AGB dust has an extremely low Fe content, which is consistent with our modelling, suggesting the absence of FeSi condensation due to kinetic constraints.

Similarly, the mineralogy and composition of presolar grains provide strong clues about their birth environment. It is commonly considered that the majority of these grains come from C-rich AGB with abundant SiC[47] and rare C[48], CaS and MgS, and TiC (but also (Zr, Mo)C has also been documented as tiny grains[49–52]). In contrast, the presence of FeSi is very rare and is only mentioned as a small inclusion in ref. 52. Thus, to the first order, our modelling is consistent with the observations in presolar grains. In some cases, TiC has been described as occurring as inclusions inside SiC or graphite dust grains, and previous studies have suggested that TiC (or ZrC) could have acted as a seed for heterogeneous nucleation[50,53–57].

## Conclusion

Overall, our experiment validated the mineralogical sequence that had been predicted by equilibrium thermodynamic calculations and showed that under kinetic conditions where condensation is limited by flow rate, the equilibrium sequence of minerals is preserved, whereas the phase abundance is under kinetic control. By including the effect of light interaction with minerals that have strong infrared absorption, our results also showed that one can effectively predict the observed minerals found in C-rich stars as well as in presolar grains. However, it should be mentioned that the wind characteristics of AGB stars could be notably affected in the case of binary systems with closed-by companions, which often have equatorial density enhancement, accretion disk and complex inner wind dynamics[58]. Nevertheless, this study suggests that more extensive condensation experiments, such as that presented here, could provide new insights into dust formation in astrophysical environments.

## Methods

### Experimental approach

The experiment was performed at Mines ParisTech-PERSEE, Sophia Antipolis, Nice, France (Fig. 1). The experimental setup consisted of a large-volume plasma torch operating at atmospheric pressure. Finely ground powder (~20 μm) of the NWA 869 ordinary chondrite (bulk composition given in Extended Data Table 1) was injected into a 5,000 K Ar–H$_2$ plasma formed between three graphite electrodes. The total volume of plasma thus formed was about 1 L. Electrode erosion enriched the vapour with carbon, resulting in a constant C/O ratio near unity. This ratio was substantially higher than the initial C/O ratio of the starting chondritic material. The vapour formed in the plasma flowed downwards in the graphite-coated chamber, along which condensation occurred in a temperature gradient. A removable T-shaped graphite rod positioned at the centre of the chamber facilitated the retrieval of condensates formed due to the thermal gradient of the chamber. The temperature and flow velocity were numerically determined. The magnetohydrodynamic equations of the gas and the thermomechanical evolution of 2,000 virtual particles were solved using the ANSYS Fluent software[26]. Thus, we were able to compute the temperature field inside the condensation chamber (Fig. 1b) and the gas flow streamlines. The flow velocity ranged between 1 and 0.6 m s$^{-1}$ at the top and end of the chamber, respectively, whereas the temperature ranged from 2,400 to 1,300 K at the top and end of the graphite rod, respectively. The rod was recovered after the experiment and subjected to chemical and mineralogical analyses. Condensates on the T-shaped graphite rod were collected using a scraping technique from 14 segments, each 5 cm long, along their lengths (Extended Data Figs. 2 and 3). The chemical and mineralogical compositions of each segment were then analysed and are summarized in Extended Data Tables 1 and 2, respectively.

**Thermodynamic simulations**

Thermodynamic simulations of the system were performed using FactSage software. This software computes the equilibrium composition of a system based on the Gibbs energy minimization method, given the composition and $P$–$T$ conditions of the system. We used the FactPS, FTStel and FToxid databases[59]. By calculating the equilibrium phase composition as a function of decreasing temperature, the software can simulate condensation starting from a vapour with a chondritic composition, mixed with Ar, $N_2$, $H_2$ and additional C.

The system was composed of 600 g of the NWA 869 composition (Extended Data Table 1) and an amount of Ar, $H_2$ and $N_2$ that corresponded to the total amount of gas flowing in the plasma torch during the experiment. As the actual C/O ratio of the system could not be easily determined, several simulations were performed with C/O molar ratios ranging from 0.72 to 1.15. For each simulated composition, the pressure was maintained at a constant value of $10^5$ Pa, whereas the temperature ranged between 2,700 and 1,300 K, with temperature steps of 50 K.

For each temperature step, the software computed the expected phases at equilibrium, including their composition and total mass. We considered three possible condensation behaviours: an equilibrium condensation in which the bulk composition of the system remained constant along its path, a fractional condensation where all the condensed material was subtracted from the vapour composition at each temperature step and a partially fractional condensation where only a fraction $f$ of what was supposed to condense at equilibrium could effectively condense due to kinetic effects (see the next section). For each of these scenarios, we were able to define the condensation sequence and bulk chemical pattern of the condensates with decreasing temperature. These results were then compared with the actual sequence and the chemical pattern measured on the T-shaped rod.

**Thermo-kinetic model of condensation**

Because the data could not be explained by an equilibrium model or a fractional condensation model, we designed a kinetic model that considers the kinetics of condensation as vapour flows through the condensation chamber. The schematic geometry of the experiment is presented in Extended Data Fig. 3. The system was composed of the condensation chamber with radius $r$ = 7 cm. At the centre of the chamber, the rod was assumed to be a cylinder with radius $r'$ = 1 cm. Only a fraction $f$ of what was expected to condense in equilibrium could effectively condense due to kinetic limitations. To evaluate the value of $f$ that would be consistent with the kinetic constraints, we included a model for gas flow through the graphite-coated chamber. We considered that the kinetics of condensation were controlled by the gas flow, thus neglecting the kinetics of nucleation. The vapour was assumed to flow downwards through a vertical cylinder of diameter $r$. If one considers a slice of thickness d$z$ through a cylindrical tank located below the plasma, one can write a mass balance equation for the fluxes (Extended Data Fig. 4):

$$F_{\text{in}}(z) = F_{\text{out}}(z + dz) + F_{\text{cond}}(z) \tag{1}$$

where $F_{\text{in}}$ is the advective flux of gas entering at position $z$, $F_{\text{out}}$ is the advective flux of gas leaving at position $z + dz$ and $F_{\text{cond}}$ is the condensation flux on the inner surface of the chamber.

One can then write $F_{\text{in}}$ and $F_{\text{out}}$ as a function of the molecular densities $\rho(z)$ and $\rho(z + dz)$, respectively:

$$F_{\text{in}}(z) = \rho(z) S v$$

$$F_{\text{out}}(z + dz) = \rho(z + dz) S v$$

$S$ is the surface of an axial section through the system, that is, $S = 2\pi(r^2 - r'^2)$; $v$ is the mean vertical gas velocity. The condensation of elements follows the Hertz–Knudsen equation:

$$F_{\text{cond}}(z) = \frac{S' \alpha (P_i(z) - P_{i,\text{sat}}(z))}{\sqrt{2\pi M R T(z)}}$$

$S'$, the total condensation surface, is composed of the rod and chamber sides; $M$ is the molar mass of the evaporated species; and $\alpha$ is the condensation coefficient (or sticking coefficient). For a small slice of length d$z$, the total condensation surface is $S' = 2\pi(r + r')dz$.

Considering the vapour phase as an ideal gas, one can write $\rho_i(z) = \frac{P_i(z)}{RT(z)}$. The flux conservation equation thus becomes

$$\rho_i(z) S v = S' \frac{\alpha(P_i(z) - P_{i,\text{sat}}(z))}{\sqrt{2\pi M R T(z)}} + \rho_i(z + dz) S v$$

This equation can be further transformed as follows:

$$\left(\frac{P_i(z)}{RT(z)} - \frac{P_i(z + dz)}{RT(z + dz)}\right) S v = \frac{S' \alpha (P_i(z) - P_{i,\text{sat}}(z))}{\sqrt{2\pi R T(z)}}$$

$$\frac{d\left(\frac{P_i}{T}\right)}{dz} dz = \frac{2\pi(r + r') dz\, \alpha (P_i(z) - P_{i,\text{sat}}(z + dz))}{S v} \sqrt{\frac{R}{2\pi M T(z)}}$$

$$\frac{T(z) \frac{dP_i}{dz} - P_i \frac{dT}{dz}}{T^2(z)} = \frac{2\pi(r + r') \alpha (P_i(z) - P_{i,\text{sat}}(z))}{S v} \sqrt{\frac{R}{2\pi M T(z)}}$$

The final equation is then written and can be integrated numerically:

$$\frac{dP_i}{dz} = \frac{2\pi(r + r') \alpha (P_i(z) - P_{i,\text{sat}}(z))}{S v} \sqrt{\frac{RT(z)}{2\pi M}} + \frac{P_i}{T(z)} \frac{dT}{dz} \tag{2}$$

The first term on the right-hand side of the equation corresponds to the condensation flux, and the second term is the pressure variation linked to the temperature gradient d$T$/d$z$, which was computed using the calculated temperature from Fluent software. The saturation pressures for the different species in the gas phase were evaluated using the equilibrium values calculated using FactSage.

Equation (2) was integrated numerically to calculate $P_i(z)$ based on the computed thermal gradient and vapour pressure $P_{i,\text{sat}}$, which was calculated in the thermodynamic simulations described previously.

For the sake of simplicity, we focused on elements with a reduced number of species in the vapour phase, which condensed in a reduced number of phases, such as $Fe_{(g)}$, which condensed in the FeSi molten alloy. We also applied this model to $Si_{(g)}$ in the upper part of the rod because it condenses in the FeSi alloy as well, and $Ca_{(g)}$, which condenses only in the oxide melt and oldhamite when $T > 1{,}600$ K. The fraction of element $i$ that effectively condensed relative to the maximum amount that would have condensed, had we reached chemical equilibrium, can be expressed as

$$f_i = \frac{n_i(T) - n_i(T + \Delta T)}{n_i(T) - \overline{n_{i,\text{sat}}}(T)} \quad (3)$$

where $n_i(T)$ is the number of moles of $i$ at temperature $T$, and $\overline{n_{i,\text{sat}}}(T)$ is the mean number of moles of $i$ in equilibrium over the $[T, T + \Delta T]$ interval with $\Delta T = 50$ K. Our thermodynamic calculations show that $\overline{n_{i,\text{sat}}}(T) \ll n_i(T)$ at all temperatures in the experiment. Thus, the value of $f_i$ can be approximated with

$$f_i \approx \frac{n_i(T) - n_i(T + \Delta T)}{n_i(T)} \quad (4)$$

The $f_i$ value depends implicitly on the oversaturation of a species $P_i/P_{(\text{sat},i)}$ in the vapour and on the condensation coefficient $\alpha$, which is a critical parameter for computing the value of $f_i$. The condensation coefficient of iron was determined experimentally[19] at pressures lower than those in our experiment and reported a value of 1 for oversaturation greater than 10. However, for other elements, it has been shown that the condensation coefficient decreases with pressure. The measured chemical pattern of Fe (Fig. 4) is best fit by an $f$ value of 6–8%, corresponding to an $\alpha$ value of 0.4, which is within the range of condensation coefficients reported in ref. [19]. For calcium, a value of 0.2 for $\alpha_{Ca}$ yielded the best fit, which is consistent with published data for Ca evaporation[60,61] and with the mean value for divalent cation evaporation from silicate melt[62,63], especially in H-rich environment[60]. The $\alpha_{Si}$ was poorly constrained in a previous study, and its value was adjusted to 0.45. The values of the condensation factor $f$ are reported in Extended Data Table 3 for various types of condensates. The best fit was obtained with a constant condensation fraction for all chemical species in a given phase. For all elements, we found $f_i$ values well below 1 (Extended Data Table 3), expressing the kinetic limitation in the condensation rate due to the effect of gas flow and kinetic barrier represented by the condensation coefficient $\alpha_i$.

As shown in Extended Data Fig. 4, the condensation fraction $f$ of Fe assessed with this model shows a moderate increase from 6% at 2,280 K to 8.2% at 1,377 K with a mean value of 7%. The computed $f_{Si}$ value shows the same pattern as the temperature, with a mean value of 8.3% in the region where Si condenses in the FeSi metal. For Ca, two regime stages of condensation were considered: the first region where Ca condenses only in the oxide melt ($T > 1{,}750$ K) and the second a region where Ca condenses both in the silicate melt and in the sulfide phase ($T < 1{,}750$ K). This leads to a concave curve with an increase in $f_{Ca}$ from 1.5% at 1,950 K to 2.6% at 1,618 K. Finally, the $f$ factors of the different phases used for the thermodynamic calculations are presented in Extended Data Table 3.

### Optical properties and condensation distances

The condensation of dust in the outflow of C-rich stars should, to some extent, be controlled by the thermodynamics of solid-phase formation with a vapour composition at a high C/O ratio. However, it has been shown in a previous study[64] that the radiative properties of condensed dust play a critical role in addition to its thermodynamic properties. The photons radiated by the star interact with the dust: a fraction of the light is reflected, another fraction is absorbed and the last fraction is transmitted. If the fraction of absorbed light is notable, it will contribute to heating the dust to a temperature that could be higher than the temperature of the ambient gas. In turn, this higher temperature could delay dust condensation. This effect contributes to an increase in the oversaturation of this phase. As the gas and dust flow away from the star, the temperature and pressure decrease, such that at some point, the predicted dust temperature corresponds to its condensation temperature[38,39]. At this radial distance, the dust can condensate.

In detail, the fraction of absorbed light depends critically on the optical parameters of the dust, and it has been shown in numerous studies that this depends on the Fe content, among other parameters. When the Fe content increases, the absorption coefficient increases, especially in the frequency domain, where radiation is efficient at heating the material. In the case of small dust particles (micrometre scale), the parameter used to describe light absorption is $Q_{\text{abs}}$, and by plotting $Q_{\text{abs}}/a$, where $a$ is the diameter of the grain, as a function of photon energy, one can define a parameter $p$ that is then used to calculate the dust temperature $T_d$ (ref. [39]). The parameter $p$ was calculated for several oxides and amorphous carbon, but not for the mineral phases that are predicted to condense in a high C/O environment. Based on the published optical properties of FeSi, TiC, AlN, SiC and graphite, we calculated the value of parameter $p$ to predict the corresponding dust temperature (Extended Data Table 4).

The study in ref. [39] has shown that the distance of condensation of a particular mineral species can be calculated based on the value of parameter $p$ and the temperature at the surface of the star $T^*$:

$$T_d(r) = T^* \left(\frac{2r}{R^*}\right)^{-\frac{2}{4+p}} \quad (5)$$

If the condensation temperature $T_c$ is known independently from thermodynamic calculations, it is then possible to determine the radial distance for condensation of a particular mineral phase by setting

$$T_d(r) = T_c(r) \quad (6)$$

For the purpose of our calculations, we have used the $T(r)$ and $P(r)$ curves for a C-rich star given in ref. [14], and the results of our condensation calculations are presented graphically in Fig. 6.

### Calculations of parameter $p$ for mineral phases condensed at a high C/O ratio

The value of $p$ was calculated for the following mineral phases based on the published optical properties: TiC, AlN, CaS and FeSi (Extended Data Table 4). In the case of FeSi, the value of $\alpha_{\text{abs}}$, the absorption coefficient, was calculated using the following equation, which depends on the bandgap energy[65]:

$$\alpha_{\text{abs}} = \frac{A}{\omega}\sqrt{\omega - E_g}$$

where $E_g$ is the bandgap energy. In the case of FeSi, which is a semiconductor at low temperatures but a metal at high temperatures, the dependence of $E_g$ on temperature was taken into account using the following equation[66]:

$$E_g = E_{g0} - S\hbar\omega\left(\coth\left(\frac{\hbar\omega}{2kT}\right) - 1\right)$$

The value of $k$, the imaginary part of the refraction index, was then calculated as

$$k(\lambda, T) = \frac{\alpha_{\text{abs}}\lambda}{4\pi}$$

Then, the value of the reflectivity given in ref. 67 was used to calculate the value of the real part of the complex refraction index ($n + ik$) using the following equation:

$$R = \frac{(n-1)^2 + k^2}{(n+1)^2 + k^2}$$

Finally, the absorption parameter $Q_{abs}$ was calculated using the following expression[64]:

$$Q_{abs} = 4\frac{2\pi a}{\lambda}\frac{6nk}{(n^2 - k^2 + 2)^2 + 4n^2k^2}$$

In the case of CaS, experimental studies have reported values of the real and imaginary dielectric functions as a function of energy[68]. These parameters are directly related to the real $\varepsilon_1$ and imaginary $\varepsilon_2$ parts of the complex refraction index:

$$\varepsilon_1 = n(\omega)^2 - k(\omega)^2$$

$$\varepsilon_2 = 2n(\omega)k(\omega)$$

From these equations, one can calculate:

$$k(\omega) = \sqrt{\frac{\sqrt{\varepsilon_1^2 + \varepsilon_2^2} - \varepsilon_1}{2}}$$

$$n(\omega) = \sqrt{\frac{\sqrt{\varepsilon_1^2 + \varepsilon_2^2} + \varepsilon_1}{2}}$$

The equation given previously was then used to calculate $Q_{abs}$. In the case of TiC, the absorbance was directly determined by ab initio calculation[69]. For AlN, the value of $Q_{abs}$ is a function of frequency that was directly calculated from the equation based on $n$ and $k$, as given in ref. 70. Optical parameters used for the calculation (Extended Data Table 4) were taken from references[38,67–73].

### Modelling the timescale of grain growth relative to the timescale of outflows

The time necessary to grow a dust grain starting from a seed, such as a more refractory grain, is highly dependent on the local density of the constitutive atoms of the considered dust. If one takes the example of FeSi (Extended Data Fig. 4), the growth rate will be limited by the rate at which individual atoms hit the grain surface. In the case of dust condensation in stellar outflow, we calculated the maximum rate of condensation by neglecting the rate of evaporation. In all cases, the calculated rates should be greater than the actual rates. As detailed in refs. 33,41, the condensation rate of FeSi is limited by the rate of Si abundance because it is lower than that of Fe. Thus, one can write the timescale of grain growth as

$$\frac{da}{dt} = \frac{V_0 \alpha (P_{Si} - P_{Si,sat})}{\sqrt{2\pi m_{Si} RT}}$$

where $V_0$ is the average volume occupied by an FeSi unit.

This relationship can be further modified according to the study in ref. 41 by including the relationship between local pressure, gas density and Si abundance:

$$P_{Si} = \frac{n_{Si}}{n_H}\frac{2n_{H_2}RT}{V} = \varepsilon_{Si}\rho RT$$

To calculate the growth timescale as a function of distance from the star, one can use the equation given in ref. 33:

$$\tau_{growth} = \frac{a}{\frac{da}{dt}} = \frac{a\sqrt{2\pi m_{Si} RT}}{V_0 \alpha P_{Si}} = \frac{a\sqrt{2\pi m_{Si} RT}}{V_0 \alpha \varepsilon_{Si} \rho RT} = \frac{2a}{V_0 \alpha \varepsilon_{Si} \rho(r)}\sqrt{\frac{2\pi m_{Si}}{RT}}$$

One can also calculate the timescale associated with the stellar outflow. According to ref. 46, this parameter can be evaluated using the following equation:

$$\tau_{outflow} = \frac{2}{\Gamma - 1} r \frac{v}{v_{esc}^2}$$

where $\Gamma$ is defined as the ratio of radiative acceleration to gravitational attraction (ref. 39):

$$\Gamma = \frac{f_K L_*}{4\pi c G M_*}$$

This parameter increases outwards and is above 1, ranging between 2 and 5 (ref. 37), and $v_{esc}$ is the escape velocity:

$$v_{esc} = \sqrt{\frac{2GM_*}{4R_*}}$$

The velocity of the outflow at an infinite distance can be calculated using the following equation[37]:

$$v = \sqrt{\frac{2GM_*}{r_0}(\Gamma - 1) + v_0^2}$$

where $r_0$ represents the radius at which acceleration starts (dust condensation radius). According to ref. 37, the initial velocity at $R_*$ can be approximated as sound velocity[71,74], which can be defined as

$$v_0 = \sqrt{\frac{C_p^{H_2} RT}{M_{H_2}(C_p^{H_2} - R)}}$$

The value of $C_p$ for $H_2$ was taken from the National Institute of Standards and Technology (NIST) compilation by refs. 72–76:

$$C_p^{H_2} = 18.563083 + 12.257357 \times T - 2.859786 \times T^2$$
$$+ 0.268238 \times T^3 + \frac{1.977990}{T^2}$$

### Data availability
All data needed to evaluate the conclusions of this study are presented in this paper and in the Supplementary Information. Additional data related to this study may be requested from the authors.

## Acknowledgements
We are particularly grateful to F. Fabry (PERSEE, Mines ParisTech, Nice, France) for his assistance with running the plasma torch. We are grateful to S. Höfner for her insightful advice during the early stages of this project, and M. Chaussidon for discussions. This project was financially supported by funding from the Fédération de Recherche Wolfgang Doeblin, FR 2800, and the Université Côte d'Azur through its Academy of Complex Systems (G.L.). This study also received partial support from the ERC-funded project COSMOKEMS #694819 (B.B.).

## Author contributions
G.L. designed the study and performed the experiments with L.F., V-J.R., C.G. and F.C. G.L, C.G, V.G. and P.V. determined the mineralogy of the condensates. M.M., B.B. and G.L. reduced the data and performed the thermodynamic calculations and models. G.L., M.M. and B.B. wrote the paper under the supervision of E.L. for asymptotic giant branch implications.

## Competing interests
The authors declare no competing interests.

## Additional information
**Extended data** is available for this paper at https://doi.org/10.1038/s41550-024-02393-7.

**Supplementary information** The online version contains supplementary material available at https://doi.org/10.1038/s41550-024-02393-7.

**Correspondence and requests for materials** should be addressed to Guy Libourel.

**Peer review information** *Nature Astronomy* thanks Angela Speck and the other, anonymous, reviewer(s) for their contribution to the peer review of this work.

**Reprints and permissions information** is available at www.nature.com/reprints.

**Publisher's note** Springer Nature remains neutral with regard to jurisdictional claims in published maps and institutional affiliations.

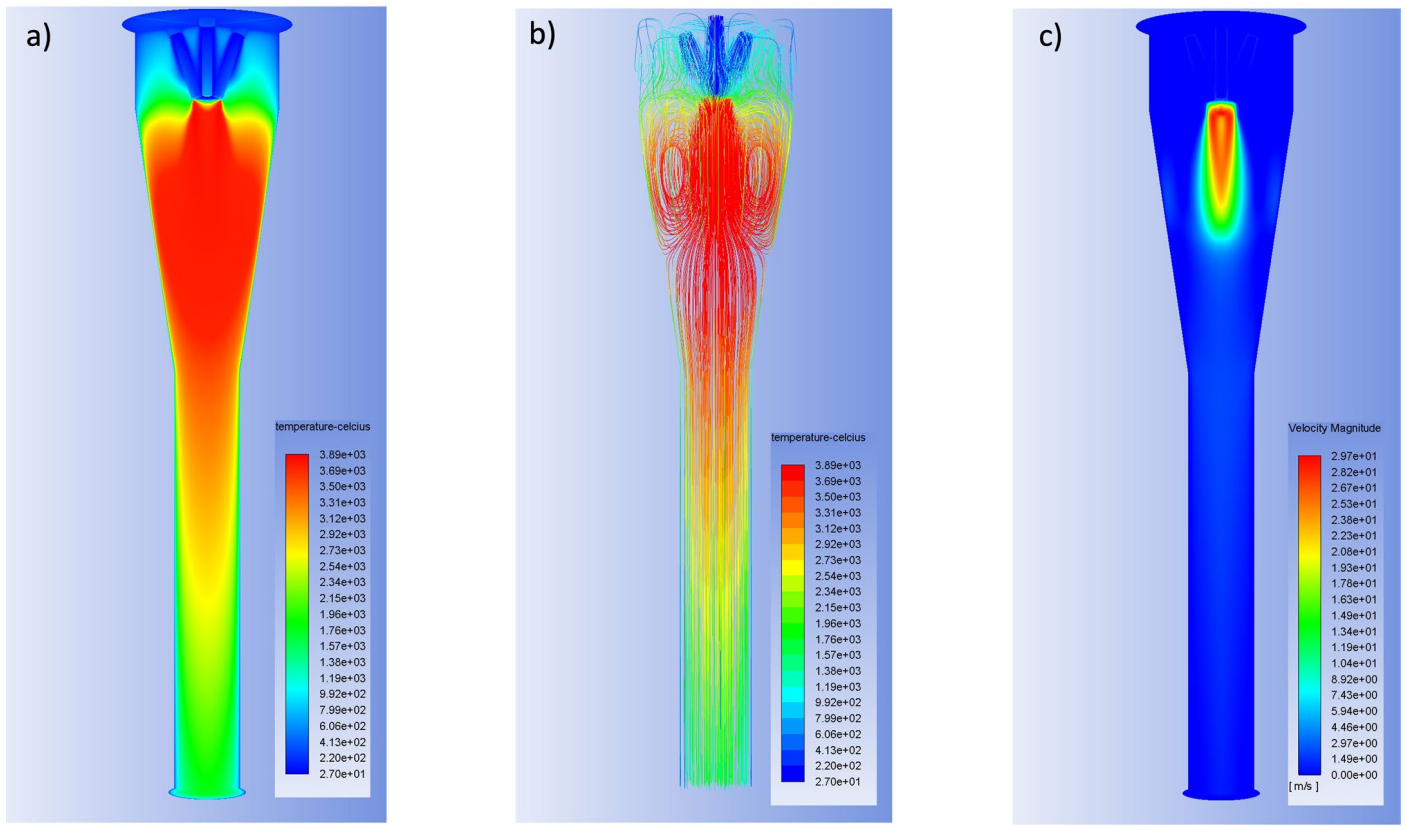

**Extended Data Fig. 1 | 2D thermal regime.** The 2D thermal regime (**a**) inside the chamber as well as the trajectory (**b**) and velocity of fictive particles (**c**) have been obtained using the ANSYS Fluent computational fluid dynamics software (see Methods).

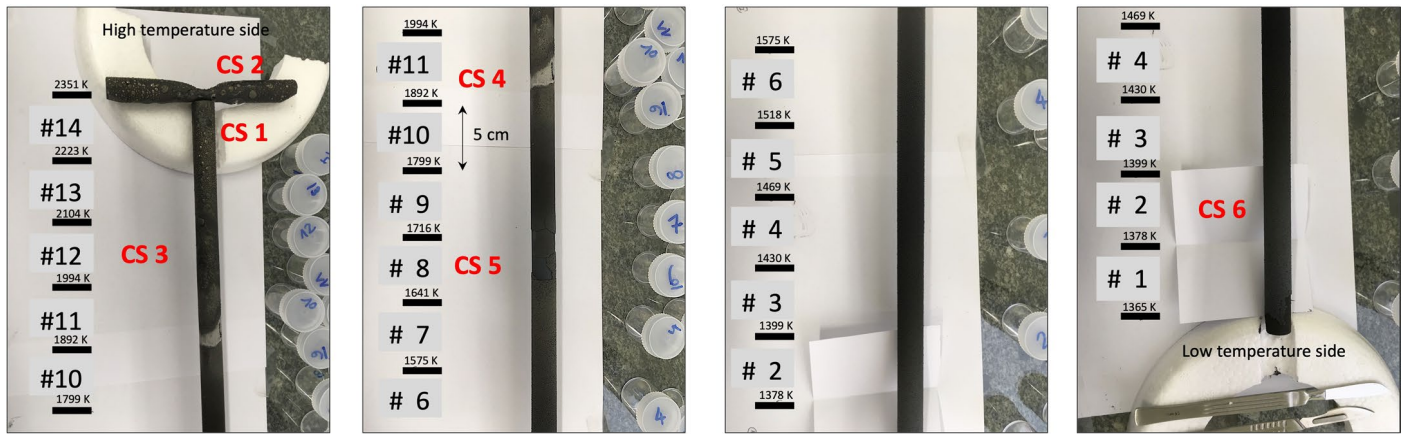

**Extended Data Fig. 2 | Samples.** Photograph of the graphite rod sample holder after the experiment that collected the condensates. As shown, 14 segments of 5 cm each have been sampled for bulk chemical analysis. CS1-CS6 correspond to sampling for detailed mineralogy depicted in Extended Data Fig. 3. See the chemistry and mineralogy of the condensates in Extended Data Table 1 and Table 2, respectively.

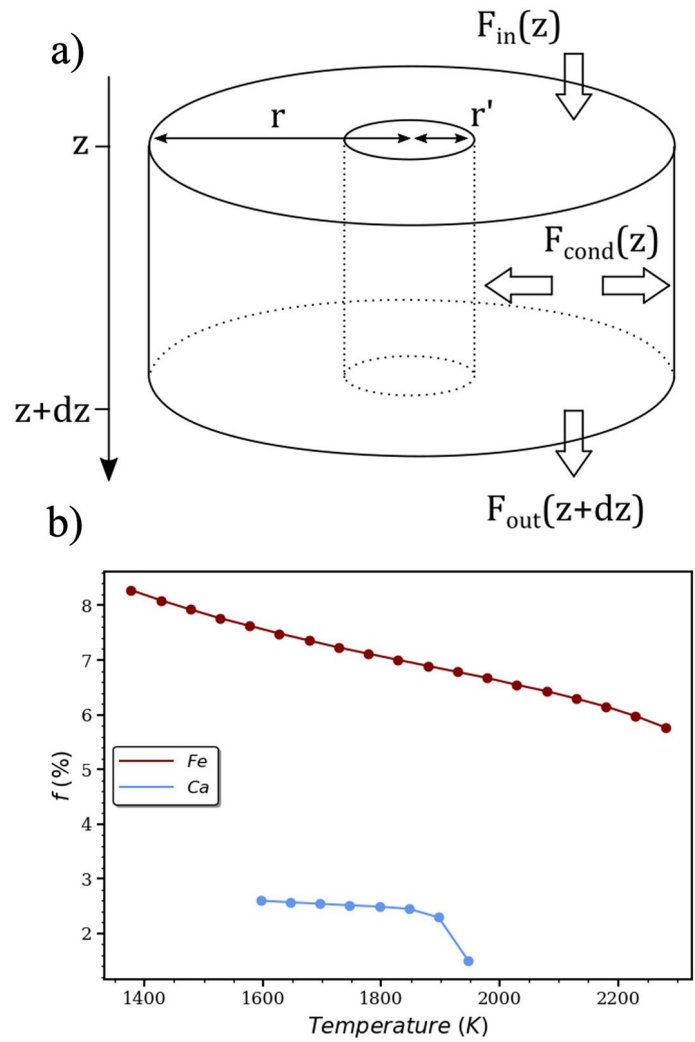

**Extended Data Fig. 3 | Kinetic model.** Schematic view of the kinetic model. **a**) The internal cylinder is the graphite rod. The total surface of evaporation corresponds to the graphite rod and the chamber's sides. **b**) Condensation factor of Ca (red) and Fe (blue). For temperatures higher than 1750 K, Ca partitions only in the oxide melt, then in oxide melt and CaS.

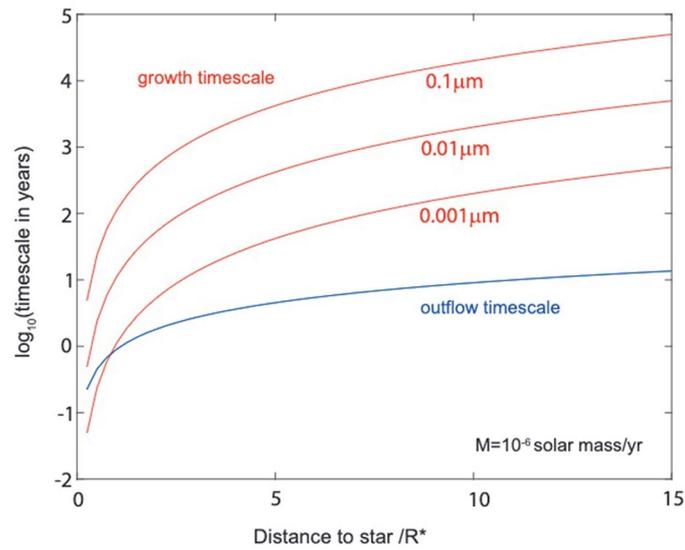

**Extended Data Fig. 4 | FeSi growth timescale.** Calculated timescales of FeSi growth (red) and stellar outflow (blue) as a function of the distance of the star (normalized to star radius $R_*$). Red curves are for Si growth for grains with a size labeled on the curve. Assuming a larger grain diameter would lead to longer timescales. Sticking coefficient of $Si_{(g)}$ is taken from ref. 76.

**Extended Data Table 1 | Chemical Composition**

| Calculated average temperature* in K | # sample from top of T-rod | Chemistry / Depth of sampling in mm | Si | Al | Fe | Mn | Mg | Ca | Na | K | Ti | P | O | C | S |
|---|---|---|---|---|---|---|---|---|---|---|---|---|---|---|---|
| | | **NWA869 chondrite bulk composition** | **19.28** | **1.20** | **19.01** | **0.28** | **15.24** | **1.54** | **0.69** | **0.10** | **0.07** | **0.11** | 43.28 | 0.12 | 2.38 |
| | | *Det. Limit* | *0.02* | *0.02* | *0.01* | *0.01* | *0.02* | *0.02* | *0.01* | *0.02* | *0.01* | *0.04* | | | |
| 2308 | #14 | 0-50 | 24.95 | 0.16 | 42.81 | 0.02 | 0.03 | 0.09 | < d.l. | < d.l. | 0.14 | < d.l. | | 26.28 ^ | 0.04 |
| 2142 | #13 | 50-100 | 26.59 | 0.24 | 44.40 | 0.02 | < d.l. | 0.07 | < d.l. | < d.l. | 0.14 | < d.l. | | 23.55 ^ | 0.04 |
| 2042 | #12 | 100-150 | 31.61 | 0.79 | 57.07 | 0.12 | < d.l. | 0.06 | < d.l. | < d.l. | 0.25 | < d.l. | | 2.59 ^ | 0.02 |
| 1949 | #11 | 150-200 | 14.43 | 1.99 | 30.40 | 0.11 | 0.33 | 1.44 | < d.l. | < d.l. | 0.07 | 0.11 | | 43.59 ^ | 0.43 |
| 1833 | #10 | 200-250 | 13.70 | 1.24 | 18.18 | 0.13 | 8.46 | 1.68 | < d.l. | < d.l. | 0.04 | 0.13 | | 38.37 ^ | 0.04 |
| 1741 | #9 | 250-300 | 23.37 | 1.42 | 20.31 | 0.30 | 18.92 | 1.86 | < d.l. | < d.l. | 0.04 | 0.21 | | 0.40 | 0.33 |
| 1685 | #8 | 300-350 | 22.67 | 1.10 | 14.62 | 0.36 | 22.27 | 1.49 | < d.l. | < d.l. | 0.03 | 0.21 | | 1.34 | 3.42 |
| 1615 | #7 | 350-400 | 18.74 | 0.84 | 11.01 | 0.34 | 22.76 | 1.16 | < d.l. | < d.l. | 0.02 | 0.18 | | 6.59 | 3.50 |
| 1556 | #6 | 400-450 | 15.97 | 0.72 | 8.82 | 0.35 | 22.12 | 1.00 | < d.l. | < d.l. | 0.02 | 0.16 | | 8.78 | 3.06 |
| 1502 | #5 | 450-500 | 15.38 | 0.70 | 8.09 | 0.39 | 22.05 | 0.98 | < d.l. | < d.l. | 0.02 | 0.15 | | 10.06 | 3.08 |
| 1453 | #4 | 500-550 | 15.64 | 0.70 | 7.76 | 0.45 | 22.57 | 0.99 | < d.l. | < d.l. | 0.02 | 0.15 | | 10.74 | 3.11 |
| 1395 | #3 | 550-600 | 15.88 | 0.71 | 7.74 | 0.51 | 24.32 | 1.01 | 0.02 | < d.l. | 0.02 | 0.14 | | 10.24 | 3.08 |
| 1389 | #2 | 600-650 | 16.36 | 0.73 | 7.87 | 0.52 | 24.81 | 1.05 | 0.03 | < d.l. | 0.02 | 0.15 | | 9.08 | 3.35 |
| 1365 | #1 | 650-700 | - | - | - | - | - | - | - | - | - | - | | 8.23 | 3.08 |

Chemical compositions in wt % of the chondrite starting material and of the bulk composition of sampled condensates shown in Extended Data Fig. 1.

**Extended Data Table 2 | Mineralogy**

| Calculated average temperature* in K | # sample from top of T-rod | Depth of sampling in mm | Mineralogy | C | FeSi | SiC | TiC | Al₄SiC₄ | CaS | MgS | AlN | TiN | Si-glass | MgAl₂O₄ | Mg₂SiO₄ | MgO | MgSiO₃ |
|---|---|---|---|---|---|---|---|---|---|---|---|---|---|---|---|---|---|
| 2308 | #14 | 0-50 |  | - | ✓ | ✓-6H | ✓ | ✓ |  |  |  |  |  |  |  |  |  |
| 2142 | #13 | 50-100 |  | - | ✓ | ✓-6H |  |  |  |  |  |  |  |  |  |  |  |
| 2042 | #12 | 100-150 |  | - | ✓ | ✓-6H |  |  | ✓ |  | ✓ |  | ✓ (Al-rich) |  |  |  |  |
| 1949 | #11 | 150-200 |  | - | ✓ | ✓-6H |  |  | ✓ |  | ✓ | ✓ | ✓ |  | ✓ |  |  |
| 1833 | #10 | 200-250 |  | - | ✓ | ✓-6H |  |  |  |  |  |  | ✓ | ✓ | ✓ |  |  |
| 1741 | #9 | 250-300 |  | ✓ | ✓ |  |  |  |  |  |  |  | ✓ |  | ✓ |  |  |
| 1685 | #8 | 300-350 |  | ✓ | ✓ |  |  |  | ✓ | ✓ |  |  | ✓ |  | ✓ | ✓ | (✓) |
| 1615 | #7 | 350-400 |  | ✓ | ✓ |  |  |  |  |  |  |  | ✓ |  | ✓ | ✓ |  |
| 1556 | #6 | 400-450 |  | ✓ | ✓ |  |  |  |  |  |  |  | ✓ |  | ✓ | ✓ |  |
| 1502 | #5 | 450-500 |  | ✓ | ✓ |  |  |  |  |  |  |  | ✓ |  | ✓ | ✓ | ✓ |
| 1453 | #4 | 500-550 |  | ✓ | ✓ |  |  |  |  |  |  |  | ✓ |  | ✓ | ✓ | ✓ |
| 1395 | #3 | 550-600 |  | ✓ | ✓ |  |  |  |  |  |  |  | ✓ |  | ✓ | ✓ | ✓ |
| 1389 | #2 | 600-650 |  | ✓ | ✓ |  |  |  |  |  |  |  | ✓ |  | ✓ | ✓ | ✓ |
| 1365 | #1 | 650-700 |  | ✓ | ✓ |  |  |  |  |  |  |  | ✓ |  | ✓ | ✓ | ✓ |

Mineralogy of the condensates sampled on the T-shape graphite rod holder as determined by Scanning Electron Microscopy and X-Ray diffraction. Further illustrations of these condensates are presented in Supplementary Figs. 1–4.

**Extended Data Table 3 | Bulk Chemistry**

| Element | mass (g) | Phase | f % |
|---|---|---|---|
| Fe | 111.5 | FeSi | 8 |
| Mn | 1.617 | SiC | 3 |
| Ti | 0.390 | AlN | 5 |
| Ca | 9.045 | | |
| K | 0.575 | CaS | 2 |
| Ar | 7132 | Silicate melt | 4 |
| S | 13.96 | Aluminates | 2 |
| P | 0.648 | | |
| Si | 113.1 | Olivine | 8 |
| Al | 7.061 | MgO | 5 |
| Mg | 89.41 | | |
| Na | 4.061 | TiC | 5 |
| O | 248.63 | Graphite | 8 |
| N | 150 | | |
| C | 173.58 | | |
| H | 179.94 | | |

Bulk composition of the starting material and values of condensation factors used in the modeling. The composition of the experiment corresponds to a C/O of 0.93 and calculated considering that 600 g of NWA869 ordinary chondrite were introduced with 4 Nm$^3$ of Ar and 2 Nm$^3$ of H$_2$.

**Extended Data Table 4 | Optical Parameter**

| Phase | p | References |
|---|---|---|
| Graphite | 2 | 76 |
| SiC | 0.3 | 77 |
| TiC | 0.8 | 71 |
| FeSi | 2.2 | 68 |
| AlN | -0.35 | 72 |
| CaS | 0.82 | 70 |
| $Mg_2SiO_4$ | -0.9 | 78 |
| $MgSiO_3$ | -0.5 | 39 |

Optical parameter p of various mineral phases.